\newcommand{\tindex}[1]{{\scriptstyle {\rm#1}}}
\documentstyle[pre,aps,epsf,floats]{revtex}

\begin{document} 
\flushbottom
\draft

\twocolumn[\hsize\textwidth\columnwidth\hsize\csname @twocolumnfalse\endcsname
  
\title{
  Depinning transition and thermal fluctuations
  in the random-field Ising model  
  } 

\author{
  L.~Roters\cite{Lars_seine_Email},
  A.~Hucht\cite{Fred_seine_Email},
  S.~L\"ubeck\cite{Sven_seine_Email},
  U.~Nowak\cite{Uli_seine_Email},
  and
  K.\,D.~Usadel\cite{Usadel_seine_Email} }
\address{
  Theoretische Tieftemperaturphysik, 
  Gerhard-Mercator-Universit\"at Duisburg,\\ 
  47048 Duisburg, Germany \\
  }
\date\today
\maketitle
\begin{abstract}
  We analyze the depinning transition of a driven interface in the
  3$d$~random-field~Ising~model (RFIM) with quenched disorder by
  means of Monte Carlo simulations. 
  The interface initially built into the system is perpendicular to 
  the \protect{$[111]$}-direction of a simple cubic lattice.
  We introduce an algorithm which is capable of simulating such
  an interface independent of the considered dimension and time scale. 
  This algorithm is applied to the 3$d$-RFIM to study both the
  depinning transition and the influence of thermal fluctuations on
  this transition.  
  It turns out that in the RFIM characteristics of the depinning
  transition depend crucially on the existence of overhangs. 
  Our analysis yields critical exponents of the interface velocity,  
  the correlation length, and the thermal rounding of the transition.
  We find numerical evidence for a scaling relation for these
  exponents and the dimension $d$ of the system.  
\end{abstract}
\pacs{68.35.Rh,75.10.Hk,75.40.Mg}

]

\section{Introduction}
\label{sec:intro}

Driven interfaces in systems with quenched disorder display with
increasing driving force a transition
from a phase where no interface motion takes place to a phase 
with a finite interface velocity. 
This so-called depinning transition is caused by a competition of 
driving force and quenched disorder.
While the driving force tends to move the interface, the motion is
hindered by the disorder (see e.g.~\cite{LNST_REV}). 

Depinning transitions are found in a large variety of physical
problems, like fluid invasion in porous media~\cite{CIEPLAK},
depinning of charge density waves~\cite{FISHER_01,FISHER_02} or
field-driven motion of domain walls in ferromagnets~\cite{BRUINSMA}.  
In magnetic systems a domain wall separates regions of different spin
orientations.   
With the assumption that the corresponding interface shows properties
of an  elastic membrane, it has been argued~\cite{BRUINSMA} that the
depinning of the interface can be described by an
Edwards-Wilkinson equation~\cite{EW} with quenched disorder. 
While the interface motion in a system with quenched disorder near the
critical threshold is theoretically often investigated in the absence
of thermal fluctuations, these fluctuations affect the experimental
study of the depinning transition~\cite{EXP1,EXP2,EXP3}.  
The crucial point is that energy barriers which are responsible for a
trapping of the interface in a metastable state at zero temperature
can always be overcome due to thermal fluctuations.
For driving fields far below the transition field this yields a
thermally activated creep motion
(see~\cite{EXP3} and references therein).   
This behavior changes approaching the transition point, where finite 
temperatures cause a rounded depinning transition (for experimental
evidence see, for instance, Fig.~2 in~\cite{EXP3}).   
To describe the dependence of the interface velocity on driving
force and temperature near the transition point, a scaling ansatz has
been proposed~\cite{FISHER_02}.
This ansatz which is based on an equation of motion for sliding
charge density waves predicts its characteristic velocity to be a
power law of temperature at the critical threshold. 
This scaling ansatz has been shown to be a valid description for the
depinning of a domain wall in the 2$d$ random-field Ising
model (RFIM) with quenched disorder~\cite{ULI_1}.

The outline of our paper is as follows:
Sec.~\ref{sec:m_and_a} describes the RFIM and reflects properties of
$[111]$-interfaces in this model.  
In Sec.~\ref{sec:Teq0} we discuss the depinning transition from a
microscopic point of view, analyzing the mechanisms of interface
motion near the depinning transition. 
Also, we determine numerically the exponents of the interface velocity
and of the correlation length, allowing an estimation of the
universality class of the 3$d$-RFIM.  
In Sec.~\ref{sec:Tl0} we analyse the influence of temperature on the
depinning transition.
By assuming the interface velocity to be a generalized homogenous
function, our analysis is based on applying standard concepts of
critical equilibrium phenomena. 
The ansatz allows the characterization of the thermal rounding of the
depinning transition by a critical exponent $\delta$. 
We determine $\delta$ for the depinning transition in the 3$d$-RFIM
for the first time and find numerical evidence for a scaling relation
among certain critical exponents characterizing this transition.
This scaling relation also holds in the 2$d$-RFIM analyzed
previously~\cite{ULI_1}. 
\section{Interfaces in the RFIM}
\label{sec:m_and_a}
We investigate the $3d$-RFIM with quenched disorder on a simple cubic
lattice.      
The Hamiltonian of the system is given by 
\begin{equation}
  {\cal H} =
  - \frac{J}{2} \sum_{\langle i,j\rangle} S_i \, S_j
  - H \sum_iS_i
  - \sum_i h_i \, S_i \,,
  \label{eq:hamiltonian}
\end{equation}
where the first sum is restricted to nearest-neighbors.  
$H$ denotes the driving field and $h_i$ quenched random-fields
which are uniformly distributed within an interval
$[-\Delta, \Delta ]$. 
We carry out Monte Carlo simulations with single-spin-flip dynamics
and we use transition probabilities  
$p(S_i \rightarrow -S_i,T)$, where $T$ denotes the temperature,
according to a heat-bath-Algorithm
(see e.\,g.~\cite{BINDER_HERRMANN} and references therein). 
At zero temperature these transition probabilities
reduce to
\begin{equation}
  p(S_i \rightarrow -S_i,0) = 
  \left\{
    \begin{array}{ll}
      1   & : \delta {\cal H} < 0 \\
      1/2 & : \delta {\cal H} = 0 \\
      0   & : \delta {\cal H} > 0\,,
    \end{array}
  \right.
  \label{eq:w_t_eq_0}
\end{equation}
where $\delta {\cal H} = {\cal H}(-S_i) - {\cal H}(S_i)$. 
We investigate three dimensional cubic systems of linear extension
from $L=12$ to $L=162$. 

An initially flat interface is built into the system separating
regions of up- and down spins.
The applied field $H$ drives the interface.
Within the Monte-Carlo simulation spins adjacent to the interface flip
causing a movement of the interface. 
Also, nucleation may occur, i.e.~a spin initially parallel to all of
its neighbors may turn. 
Since we are interested in the scaling behavior of the interface
motion in the vicinity of the depinning transition, it is essential
that within the observation time nucleation does not occur. 
The minimum energy needed for isolated spin flips is
$2(z J-H-\Delta)$. 
As long as this quantity is large as compared to temperature, 
the time scales on which nucleation and interface motion occur are
separated, and within the observation time no nucleation takes
place~\cite{ULI_1}.
In particular, there is no need to suppress artificially nucleation or
isolated spin flips during the simulation.

\begin{figure}[b]
  \epsfxsize=8.2cm
  \hspace{1.3cm}
  \epsffile{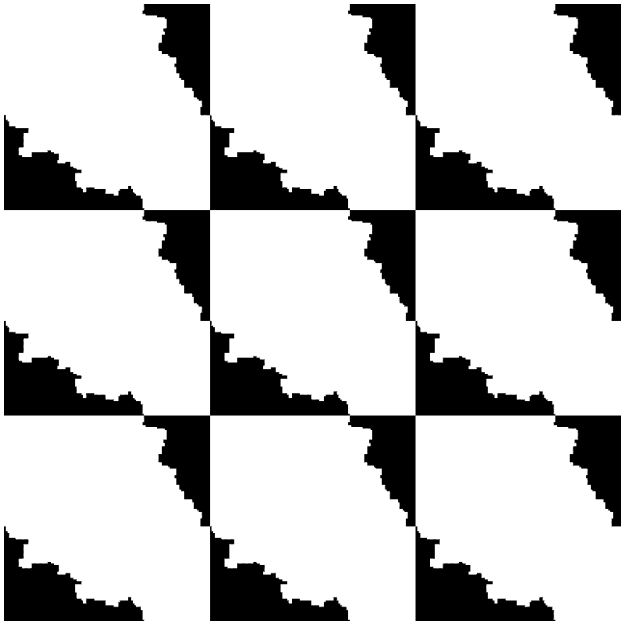}
  \vspace{0.5cm}
  \caption{
      Periodic images of a moving interface in $d=2$.
      Antiperiodic boundary conditions are applied. 
      Black areas correspond
      to $S_i<0$ and white areas to $S_i>0$.
    }
  \label{fig:interface}
\end{figure}

The analysis of interface motion on simple cubic lattices considers
usually $[100]$-interfaces. 
However, investigating $[100]$-interfaces in the limit of vanishing
disorder means that the interface motion is restricted to driving
fields $H/J > z-2$~(see \cite{ROBBINS_1}).  
To avoid this, we consider $[111]$-interfaces which move in the
absence of disorder at arbitrarily small driving
fields increasing the separation of time scales for interface motion
and nucleation even further~\cite{ULI_1}.    

We have found that the most convenient way to implement
$[111]$-interfaces in the numerics is the introduction of antiperiodic
boundary conditions. 
This implementation is illustrated in Fig.~\ref{fig:interface}. 
For simplicity, periodic images of a snapshot of an interface in $d=2$
are shown. 
As can be seen from Fig.~\ref{fig:interface}, the orientations of
\emph{up} and \emph{down} are exchanged when passing the boundaries of
the system.
Of course, the exchange of \emph{up} and \emph{down} also affects the
driving field whose sign has to be choosen in an appropriate manner. 
Our implementation will work as long as the different parts of the
interface do not interact. 
An interaction takes place if the interface width,
$w \sim L^\zeta$, is of the same magnitude as the typical distance
$a$ between two neighboring parts of the interface. 
This distance is proportional to the linear extension of the
system, $a \propto L$, independent of the considered dimension. 
Our implementation is therefore applicable to situations where
$\zeta < 1$.  
Despite this restriction antiperiodic boundary conditions have the
advantages that they can be applied to any dimension and generalized
to other orientations of the interface.
They are a natural choice for interfaces, because the moving interface
can be investigated without any time limit.
This is especially an advantage close to the depinning transition,
where the critical slowing down effect causes large relaxation
times~\cite{CSD}.
\begin{figure}[t]
  \epsfxsize=6.0cm
  \hspace{1.3cm}
  \epsffile{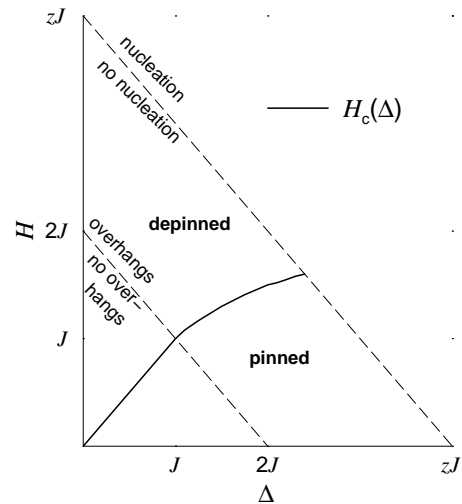}
  \vspace{0.5cm}
  \caption{
    Dependence of magnetization reversal processes on system
    parameters in the RFIM at $T=0$.  
    The dimension enters only through the number $z$ of nearest
    neighbors. 
    The bold line indicates the dependence of the critical field
    $H_\tindex{c}$ on the strength of the disorder $\Delta$.
    For $\Delta<J$ the critical field equals $\Delta$,
    while for $\Delta>J$ the bold line is only a sketch. 
    The dashed lines indicate the regions where overhangs and island
    growth appear. 
    }
  \label{fig:phase_dia} 
\end{figure}

\section{Zero Temperature}
\label{sec:Teq0}

In the RFIM with an interface initially built into the system,
there are in general two magnetization reversal processes: 
interface motion and nucleation. 
Without thermal fluctuations the second process does not occur as long
as $(H+\Delta)/J$ does not exceed the number $z$ of nearest
neighbors.
The corresponding threshold is shown in Fig.~\ref{fig:phase_dia}
(upper broken curve). 
Above this threshold nucleation processes take place and interfere
with the interface motion. 

\begin{figure}[t]
  \epsfxsize=6.8cm
  \hspace{0.5cm}
  \epsffile{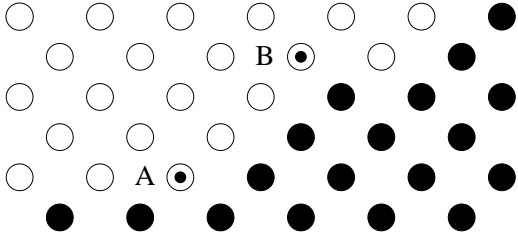}
  \vspace{0.5cm}
  \caption{
    Part of a diagonal interface in $d=2$.
    Different orientations of the spins $S_i$ are denoted by black
    circles (favored by the driving field)
    and white circles, respectively. 
    To cause a spin flip at the A, B sites, different driving fields
    are necessary (see text).  
    }
  \label{fig:domain_wall} 
\end{figure}

In the following we are interested in the influence of overhangs on
the value of the critical field $H_\tindex{c}(\Delta)$ at which
the transition takes place. 
Close to the depinning transition, there are two important
kinds of spin flips (see Fig.~\ref{fig:domain_wall}). 
All spins of type $A$ with 
\begin{equation} 
  \sum_{\langle j \rangle} S_A S_j=0
  \hspace{0.25cm}
  {\rm will~flip~if}
  \hspace{0.25cm}
  H \geq H_0 = \Delta \, ,
  \label{eq:h_0}
\end{equation}
while the first spin of type $B$ with 
\begin{equation} 
  \sum_{\langle j \rangle} S_{B} S_j=-2
  \hspace{0.25cm}
  {\rm can~flip~if}
  \hspace{0.25cm}
  H \geq H_{-2} = 2J - \Delta \, .
  \label{eq:h_m2}
\end{equation}
\begin{figure}[b]
  \epsfxsize=8cm
  \hspace{0.0cm}
  \epsffile{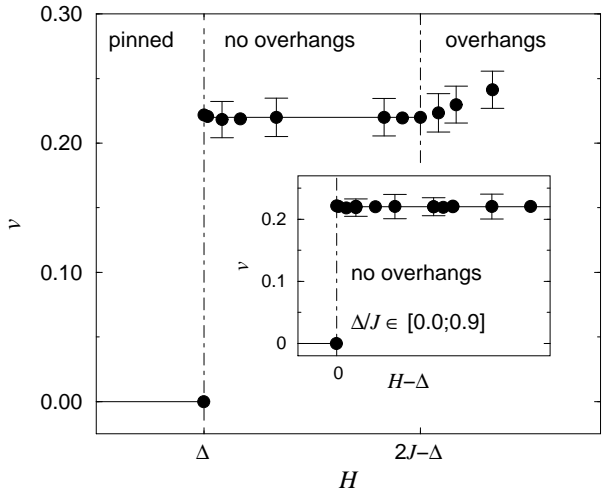}
  \caption{
    Interface velocity $v$ and its dependence on the
    driving  field $H$ for $\Delta/J=0.7$.
    The depinning transition takes place at $H=\Delta$.
    The inset shows interface velocities for different system sizes 
    $L \in \{30,42\}$ and ratios $\Delta/J<1$. 
    For reasons of clearness not all error-bars are shown.
    }
  \label{fig:delta_kleiner_j}
\end{figure}
Here, the sum is taken over nearest-neighbors of $A$ and $B$,
respectively. 
If the strength of disorder $\Delta$ is smaller than the
exchange energy $J$, then it follows that $H_0 < H_{-2}$. 
The critical field at which the transition takes place is given by
\begin{math}
  H_\tindex{c}(\Delta < J) = \Delta 
  \, .
\end{math}
Hence, no overhangs occur in the vicinity of the transition point.  
Taking into account the transition probabilities given by
Eq.~(\ref{eq:w_t_eq_0}), this value of the critical field means that
the interface velocity depends neither on the driving field nor on
the strength of disorder as long as no overhangs occur ($H<H_{-2}$). 
In particular, in the absence of overhangs the interface velocity
observed in a disordered system coincides with that of a
non-disordered system ($\Delta=0$).    
Figure~\ref{fig:delta_kleiner_j} and its inset show numerical data
which confirm this scenario for the 3$d$-RFIM within the error-bars. 

Next we investigate the depinning transition occurring in the RFIM
for $\Delta>J$. 
In this case the transition takes place at a certain field
$H_\tindex{c} < \Delta$ as can be understood from the following
consideration: For $H_0<H_{-2}$, not all spins of type $A$ can flip if
$H < \Delta$. 
But because of the existence of overhangs, a second growth mechanism
is possible to the interface:
If a spin of type $A$ cannot flip due to its large random field
$h_i$, an overhang created elsewhere can cause an avalanche by which
additional neighbors of $A$ are flipped. 
Thus the interface can be kept moving.
Contrary to the regime $\Delta<J$, the interface motion now is based
on the existence of overhangs.
Note that our considerations do not depend on the dimension $d$ of
the system, because Eqs.~(\ref{eq:h_0}) and~(\ref{eq:h_m2}) are
independent of $d$.  

We start examining the regime $\Delta > J$ in the 3$d$-RFIM
numerically by investigating the depinning transition from below. 
We analyse the disorder-averaged distance 
$\langle h(t\to\infty) \rangle$ traveled by an initially flat
interface before pinning occurs.  
This quantity is closely related to the total volume invaded by a
growing domain which was analyzed in~\cite{MARTYS,JI}. 
However, while in~\cite{MARTYS,JI} the driving force is increased
step by step to allow for relaxation processes in between, we focus
our attention to driving fields which remain unchanged during the
interface motion. 

\begin{figure}[t]
  \epsfxsize=8cm
  \hspace{0.0cm}
  \epsffile{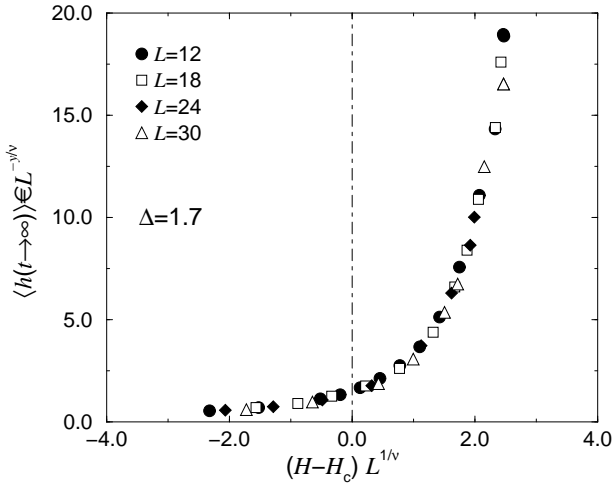}
  \caption{
    Scaling plot of the distance $\langle  h(t\to\infty )\rangle $
    traveled by the interface before pinning occurs
    as a function of the driving field $H$ according to
    Eq.~(\ref{eq:h_t_inf_scaling}).  
    The data collapse yields
    $1/\nu=1.31 \pm 0.07$,
    $y/\nu=0.98 \pm 0.4 $, and
    $H_\tindex{c}=1.371 \pm 0.003$. 
    }
  \label{fig:h_t_inf}
\end{figure}

Below the depinning transition $\langle h(t\to\infty) \rangle$ is
finite.  
Approaching the transition point with increasing driving field,
the distance traveled before pinning occurs increases and finally
diverges at the transition point. 
We assume that in the vicinity of the transition point
$\langle h(t\to\infty)\rangle $ diverges algebraically, characterized
by some exponent $y$,   
\begin{eqnarray}
  \langle  h(t\to\infty)\rangle  &
  \sim                           &  
  \left( H_\tindex{c} - H\right)^{-y} 
  \, ,
  \label{eq:h_t_inf}
\end{eqnarray}
where $H_\tindex{c}$ denotes the critical field observed in a
system of infinite extension.
In a finite system with linear dimension $L$ finite-size scaling is
assumed. The corresponding scaling ansatz reads 
\begin{equation}
  \langle  h(t\to\infty)\rangle  
  =
  L^{y / \nu} 
  \,
  f\left[ \left( H - H_\tindex{c} \right) L^{1 / \nu} \right]
  \, ,
  \label{eq:h_t_inf_scaling}
\end{equation}
with $f(x) \sim |x|^{-y}$ for $x \to -\infty$.
Note that $\langle h(t\to\infty)\rangle$ also diverges in any finite
system which means that $f(x)$ should diverge at a finite value of
$x^\star$.
The corresponding driving field defines a size dependent critical
field $H_\tindex{c}(L)$ given by
\begin{math}
  \left( H_\tindex{c}(L) - H_\tindex{c} \right) L^{1 / \nu} = x^\star
\end{math}.
A scaling plot of the data according to Eq.~(\ref{eq:h_t_inf_scaling})
is shown in Fig.~\ref{fig:h_t_inf}.
The divergence of $f(x)$ occurs at $x^\star \approx 2.5$ showing that
in a finite system the threshold field is always shifted to fields
larger that $H_\tindex{c}$.

The critical exponent of the correlation length parallel to the
interface is given by $1/\nu = 1.31 \pm 0.07$ and
the critical field turns out to be $H_\tindex{c}=1.371 \pm 0.03$.
The value of $\nu$ coincides with~\cite{JI}, where an
[100]-interface in the self-affine growth regime corresponding in our
case to $\Delta>J$ has been investigated. 
This suggests that the behavior of the correlation length
at the depinning transition does not depend on the orientation of the
interface in the RFIM.

\begin{figure}[t]
  \epsfxsize=8cm
  \hspace{0.0cm}
  \epsffile{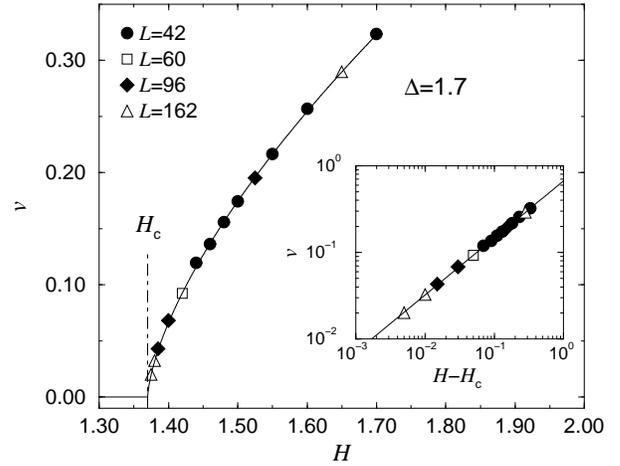}
  \caption{
    Dependence of the interface velocity $v$ on the
    driving field $H$ in the vicinity of the transition point.
    Approaching the critical field $H_\tindex{c}$,
    the system size $L$ is increased in order to avoid finite-size
    effects.
    Fitting the data according to Eq.~(\ref{eq:v_H_groesser_H_c})
    (solid line) yields
    $\beta=0.66 \pm 0.04$,
    $H_\tindex{c} = 1.37 \pm 0.01$, and
    $A = 0.67 \pm 0.03$.
    }
  \label{fig:v_H_groesser_H_c}
\end{figure}

In the following we consider the disorder averaged interface velocity
\begin{math}
  v = \langle \mbox{d}h/\mbox{d}t \rangle
\end{math}
above the transition point in the limit of large times.
This quantity can be interpreted as the order parameter of the
depinning transition.
Approaching a continuous phase transition the order parameter
vanishes in leading order according to
\begin{equation}
  v(H) = A \left(H - H_\tindex{c}\right)^{\beta}
  \,.
  \label{eq:v_H_groesser_H_c}
\end{equation}
The corresponding data are shown in Fig.~\ref{fig:v_H_groesser_H_c}.
The prefactor $A$ is a non-universal constant which can be used to
compare the results obtained at zero temperature with those
presented in the next section.
Since in the vicinity of the depinning transition finite size effects
may become important, we calculated each interface velocity $v(H)$ in 
systems of different linear extension $L$.
For sufficiently large $L$ we observed no significant dependence on
the system size from which we concluded that the data shown in 
Fig.~\ref{fig:v_H_groesser_H_c} correspond within negligible errors
to those of the limit $L \to \infty$.
As can be seen from the data, 
Eq.~(\ref{eq:v_H_groesser_H_c}) is fulfilled and we obtain
$A = 0.671 \pm 0.03$, $\beta=0.66 \pm 0.04$,
and $H_\tindex{c}=1.37 \pm 0.01$. 

The values of $\beta$ and $\nu$ obtained from our analysis
coincide within the error-bars with those of the
Edwards-Wilkinson equation with quenched disorder in $d=2+1$, 
$\beta_\tindex{EW}=2/3$ and $\nu_\tindex{EW}=3/4$.
These values are obtained by an $\epsilon$-expansion
within a functional renormalization group scheme
(see~\cite{LNST_REV,LNST}).
While the value of $\beta_\tindex{EW}$ is obtained to first order of
$\epsilon$, there are arguments that $\nu_\tindex{EW}$ is exact in all
orders to $\epsilon$~\cite{LNST_REV,NARAYAN}.  
Taking this into account, our results suggest that the depinning
transition of a domain wall in the 3$d$-RFIM with quenched disorder is
in the same universality class as the depinning transition of the
corresponding Edwards-Wilkinson equation.

\section{Finite Temperatures}
\label{sec:Tl0}

\begin{figure}[b]
  \epsfxsize=8cm
  \hspace{0.0cm}
  \epsffile{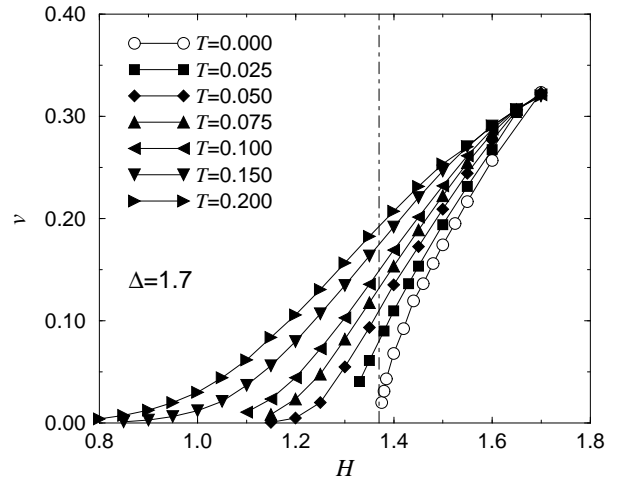}
  \caption{
    Dependence of the interface velocity on the driving field for
    different temperatures, as indicated.
    The open symbols are from 
    Fig.~\ref{fig:v_H_groesser_H_c}.
    The vertical line denotes the
    critical field obtained at $T=0$.
    }
  \label{fig:v_finite_t}
\end{figure}
In this section we study the influence of finite temperatures on the
depinning transition.  
For $T>0$ the interface velocity does not vanish for finite driving
fields since the energy needed to overcome local energy barriers is
provided by thermal fluctuations at any finite $T$.
This results in a rounded depinning transition. 
The rounding can be seen in
Fig.~\ref{fig:v_finite_t}, where interface velocities for different
driving fields and temperatures are presented. 
As expected, the rounding of the transition increases with increasing
temperature.
Again, we ensured that the interface velocities presented in this
and the following Figures correspond within negligible errors to those
of the thermodynamic limit. 
To analyse the thermal rounding of the depinning transition
quantitatively, we first note that the depinning transition can be
described in terms of a continuous non-equilibrium phase transition. 
This is suggested by the divergence of the correlation length (see
determination of $\nu$ and Fig.~\ref{fig:h_t_inf}) and
the dependence of the interface velocity on the driving field
near the transition point (Fig.~\ref{fig:v_H_groesser_H_c}).
In the standard theory of critical phenomena a continuous
phase transition is characterized by critical exponents
(see for instance~\cite{STANLEY_BOOK} and references therein). 
Beside $\beta$ describing the field dependence of the order parameter
and $\nu$ characterizing the divergence of the correlation length near the
transition point, the rounding of a phase transition is
characterized by the critical exponent $\delta$. 
In magnetic systems, for instance, $\beta$ and $\delta$ determine the
magnetic equation of state. 
We now apply this approach to the depinning transition by assuming its
order parameter to be a generalized homogenous function of temperature
and driving field,
\begin{equation}
  v\left[T,H-H_\tindex{c}\right]
  =
  \lambda \,
  v\left[\lambda^{a_T} \,T , \lambda^{a_H} \, (H-H_\tindex{c})\right]
  \, .
  \label{eq:ghf}
\end{equation}
Choosing $\lambda=T^{-1/a_T}$ we obtain the scaling ansatz
\begin{equation}
  v(T,H)
  =
  T^{1/\delta} \,
  f_T[(H-H_\tindex{c}) \, T^{-1/\beta \delta} ]
  \, ,
  \label{eq:Fisher_scaling}
\end{equation}
with $f_T(x \to 0) =\mbox{const}$. 
In particular, this equation corresponds to the magnetic equation of
state~\cite{STANLEY_BOOK}. 
From an equation of motion of sliding charge
density waves a scaling form corresponding
to Eq.~(\ref{eq:Fisher_scaling}) has been obtained~\cite{FISHER_02}. 
Note that contrary to~\cite{FISHER_02} our ansatz which is
based on Eq.~(\ref{eq:ghf}) yields no predictions on the values for
$\beta$ and $\delta$.

\begin{figure}[t]
  \epsfxsize=8cm
  \hspace{0.0cm}
  \epsffile{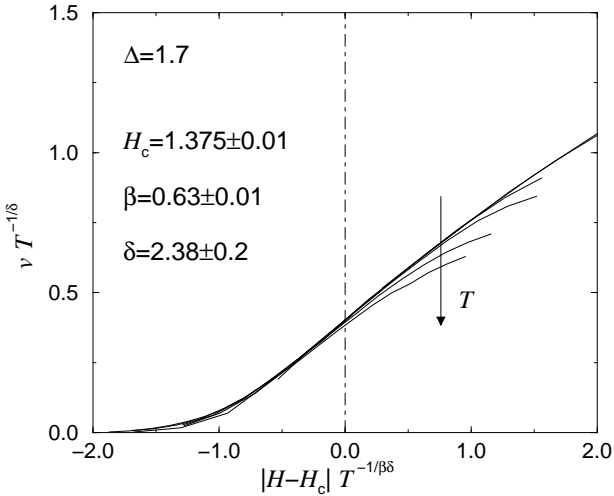}
  \caption{
    Dependence of the interface velocity on $H$ for given values of
    $T$.
    The data shown are identical to those in Fig.~\ref{fig:v_finite_t}
    for $T>0$ and rescaled according to Eq.~(\ref{eq:Fisher_scaling}).
    }
  \label{fig:delta_fisher}
\end{figure}
\begin{figure}
  \epsfxsize=8cm
  \hspace{0.0cm}
  \epsffile{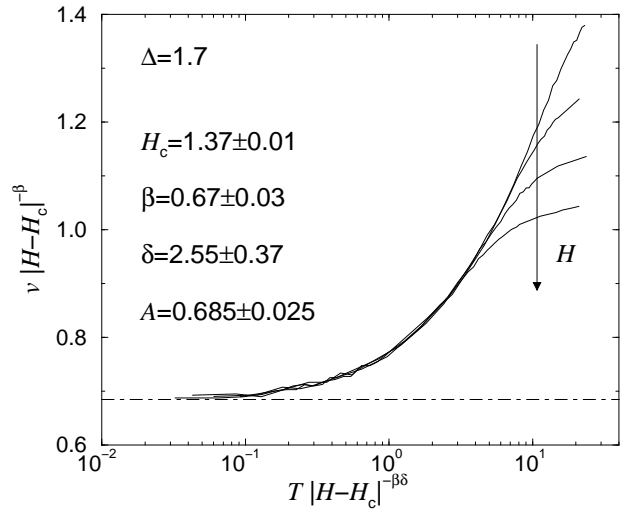}
  \caption{
    Dependence of the interface velocity on $T$ for given values of
    $H$.
    The data are rescaled according to Eq.~(\ref{eq:else_scaling}).
    The horizontal line marks the value of $A$ which is given by
    Eq.~(\ref{eq:v_H_groesser_H_c}).  
    }
  \label{fig:delta_else}
\end{figure}

It has been shown previously that Eq.~(\ref{eq:Fisher_scaling}) is
valid in the 2$d$-RFIM with quenched disorder~\cite{ULI_1}.  
We have tested this scaling ansatz in the present situation for the
3$d$-RFIM with the interface velocities shown in
Fig.~\ref{fig:v_finite_t}.   
As can be seen from Fig.~\ref{fig:delta_fisher}, the scaling ansatz
leads to a data collapse for
$\beta=0.63 \pm 0.07$, $\delta=2.38 \pm 0.2$, and
$H_\tindex{c}=1.375\pm 0.01$. 
Thus at $H=H_\tindex{c}$ the influence of temperature on the interface
velocity can be described by a power law $v \sim T^{1/\delta}$. 
To support this value for $\delta$ we can determine $\delta$ from a
different scaling function obtained from Eq.~(\ref{eq:ghf}) by
choosing $\lambda=|H-H_\tindex{c}|^{-1/a_H}$: 
\begin{equation}
  v(T,H)
  =
  (H-H_\tindex{c})^\beta \,
  f_H[ (H-H_\tindex{c})^{-\beta \delta} \, T ]
  \,,
  \label{eq:else_scaling}
\end{equation}
with $f_H(x\to 0) = \mbox{const}$. 
This ansatz is valid above the transition point and it is closely
related to Eq.~(\ref{eq:Fisher_scaling}). 
It corresponds to a different formulation of the magnetic equation of state.  
Interface velocities rescaled according to Eq.~(\ref{eq:else_scaling})
are shown in Fig.~\ref{fig:delta_else}.
One obtains
$\beta=0.67\pm 0.03$,
$\delta= 2.55 \pm 0.37$, and 
$H_\tindex{c} = 1.37 \pm 0.05$.
This result confirms within the error-bars the value of $\delta$
determined by Eq.~(\ref{eq:Fisher_scaling}). 
Beside these quantities the data collapse also allows a determination
of the prefactor $A=f_H(x \to 0)$ 
[see Eq.~(\ref{eq:v_H_groesser_H_c})] which turns out to be
$A=0.685\pm 0.025$. 

The values of $A$, $H_\tindex{c}$, and $\beta$ found for $T>0$
coincide within sufficient accuracy with those values obtained at
$T=0$.
We have demonstrated that Eqs.~(\ref{eq:Fisher_scaling}) and
(\ref{eq:else_scaling}) are valid confirming that the interface
velocity is a generalized homogenous function in the vicinity of the
transition point~\cite{STANLEY_ART}. 
Thus, the influence of temperature on the depinning transition can be
described within well-established concepts. 

The knowledge of $\beta$ and $\delta$ allows a test of the 
scaling relation $\delta = 2 + 1/\beta$ proposed by Tang and
Stepanow~\cite{TANG_STEPANOW}.
This scaling relation was shown to be fulfilled in the
2$d$-RFIM~\cite{ULI_1}. 
For $\beta \approx 0.67$ the scaling relation suggests 
$\delta \approx 3.5$ which is not supported by our results. 
On the other hand, standard theory of critical phenomena predicts
relations among critical exponents.  
For instance, combining the Rushbrooke, the Widom and the
hyperscaling relation yields in equilibrium physics 
\begin{equation}
  \delta = \frac{d \nu}{\beta} - 1
  \,.
  \label{eq:scr_2}
\end{equation}
This scaling relation is valid in dimensions $d$ below the upper
critical dimension $d_\tindex{c}$ due to the restriction of the
hyperscaling relation to $d < d_\tindex{c}$.
We have tested the scaling relation~(\ref{eq:scr_2}) with the
numerically evaluated exponents at the depinning transition and found
out that both the exponents in the present case $d=3$
as well as the exponents for $d=2$
($\nu_{2d} \approx 1.0$, $\beta_{2d} \approx 0.33$,
and $\delta_{2d} \approx 5.0$; see~\cite{ULI_1}) 
fulfill Eq.~(\ref{eq:scr_2}) within the error-bars.  
Unfortunately however, a firm foundation of this scaling relation in
the present situation for non-equilibrium phase transitions is
unknown.
\section{Conclusion}
\label{sec:con}

We investigated the motion of a driven interface in a magnetic system
with quenched disorder. 
To improve the efficiency of our numerics we applied antiperiodic
boundary conditions.
These boundary conditions allow to investigate the interface motion
on any time scale.
At zero temperature a depinning transition occurs at a finite driving
field. 
We discussed the influence of overhangs and avalanches on this
transition.  
If the strength of disorder exceeds the coupling constant, the
interface motion is based on the existence of overhangs. 
Under these circumstances the depinning transition can be
characterized by critical exponents, both below and above the critical
threshold. 
Our results suggest that the depinning transition of a domain wall in
the 3$d$-RFIM with quenched disorder and the depinning transition of
the corresponding Edwards-Wilkinson equation are in the same
universality class.   

Thermal fluctuations yield a rounded transition.
By assuming the interface velocity to be a generalized homogenous
function of temperature and driving field, this rounding can be
described within a scaling approach. 
The validity of this approach is confirmed by the fact that at the
threshold field  the interface velocity vanishes with decreasing
temperature according to a power law characterized by an exponent
$\delta$.  
We have tested a scaling relation [Eq.~(\ref{eq:scr_2})] among
different exponents characterizing the depinning transition and found
numerical evidence, that the scaling relation is valid both in the
2$d$- and the 3$d$-RFIM. 
\acknowledgments
This work was supported by the Deutsche Forschungsgemeinschaft through
the Graduiertenkolleg {\it Struktur und Dynamik heterogener Systeme}
at the University of Duisburg, Germany. 


\begin{references} 
\bibitem[*]{Lars_seine_Email}
  E-mail: lars@thp.uni-duisburg.de

\bibitem[\dagger]{Fred_seine_Email}
  E-mail: fred@thp.uni-duisburg.de

\bibitem[\ddagger]{Sven_seine_Email}
  E-mail: sven@thp.uni-duisburg.de

\bibitem[\star]{Uli_seine_Email}
  E-mail: uli@thp.uni-duisburg.de

\bibitem[\amalg]{Usadel_seine_Email}
  E-mail: usadel@thp.uni-duisburg.de




\bibitem{LNST_REV}
  H.\ Leschhorn, T.\ Nattermann, S.\ Stepanow, and L.-H.\ Tang,
  Ann.\ Physik\ {\bf 6}, 1 (1997).

\bibitem{CIEPLAK}
  M.\ Cieplak and M.\ Robbins,
  Phys.\ Rev.\ Lett.\ {\bf 60}, 2042 (1988).

\bibitem{FISHER_01}
  D.\,S.\ Fisher,
  Phys.\ Rev.\ Lett.\ {\bf 50}, 1486 (1983).

\bibitem{FISHER_02}
  D.\,S. Fisher,
  Phys.\ Rev.\ B {\bf 31}, 1396 (1985).

\bibitem{BRUINSMA}
  R.\ Bruinsma and G.\ Aeppli,
  Phys.\ Rev.\ Lett. {\bf 52}, 1547 (1984).

\bibitem{EW}
  S.\,F.\ Edwards and D.\,R.\ Wilkinson,
  Proc.\ R.\ Soc.\ London, Ser. A {\bf 381}, 17 (1982).

\bibitem{EXP1}
  F.\ Ladieu, M.\ Sanquer, and J.\,P.\ Bouchaud,
  Phys.\ Rev.\ B.\ {\bf 53}, 973 (1996).

\bibitem{EXP2}
  U.\ Nowak, J.\ Heimel, T.\ Kleinefeld, and D.\ Weller,
  Phys.\ Rev.\ B.\ {\bf 56}, 8143 (1997). 

\bibitem{EXP3}
  S.\ Lemerle, J.\ Ferr$\acute{\mbox{e}}$,
  C.\ Chappert, V.\ Mathet, and P.\ Le\ Doussal,
  Phys.\ Rev.\ Lett.\ {\bf 80}, 849 (1998). 

\bibitem{ULI_1}
  U.\ Nowak and K.\,D.\ Usadel,
  Europhys.\ Lett.\ {\bf 44}, 634 (1998).
  
\bibitem{BINDER_HERRMANN}
  K.\ Binder and D.\,W.\ Heermann,
  {\it Monte Carlo Simulation in Statistical Physics},
  Springer Series in Solid-State Sciences 80, 3rd edition,
  (Springer, Heidelberg, 1997).

\bibitem{ROBBINS_1}
  B.\ Koiller, H.\ Ji, and M.\,O.\ Robbins,
  Phys.\ Rev.\ B\ {\bf 46}, 5228 (1992).
  
\bibitem{CSD} 
  K.\ Binder, in:
  {\it Phase Transitions and Critical Phenomena},
  Vol.\ 5B,
  edited by C.\ Domb and M.\,S.\ Green,
  (Academic Press (London) Ltd., London, 1976).
  
\bibitem{MARTYS}
  N.\ Martys, M.\,O.\ Robbins, and M.\ Cieplak,
  Phys.\ Rev.\ B\ {\bf 44}, 12294 (1991).

\bibitem{JI}
  H.\ Ji and M.\,O.\ Robbins,
  Phys.\ Rev.\ B\ {\bf 46}, 14519 (1992). 

\bibitem{LNST}
  T.\ Nattermann, S.\ Stepanow, L.-H.\ Tang, and H.\ Leschhorn, 
  J.\ Phys.\ II {\bf 2}, 1483 (1992).

\bibitem{NARAYAN}
  O.\ Narayan and D.\,S.\ Fisher,
  Phys.\ Rev.\ B {\bf 48}, 7030 (1993).
 
\bibitem{STANLEY_BOOK}
  H.\,E.\ Stanley,
  {\it Introduction to Phase Transitions and Critical Phenomena},
  (Oxford University Press, Oxford, 1971).

\bibitem{STANLEY_ART}
  A.\ Hankey and H.\,E.\ Stanley,
  Phys.\ Rev.\ B {\bf 6}, 3515 (1972).
  
\bibitem{TANG_STEPANOW}
  L.-H.\ Tang and S.\ Stepanow (unpublished);
  T.\ Nattermann,
  March meeting of the American Physical Society (1993).
  
\end{references}
\end{document}